\documentclass[journal=jpccck,manuscript=article,layout=twocolumn]{achemso}
\pdfoutput=1
\usepackage[utf8]{inputenc}
\usepackage{units}
\usepackage{amsmath,amssymb,multirow,rotating}
\usepackage{graphicx,blindtext,wrapfig,lipsum}

\author{Z. Mamiyev}
\affiliation{Institut f\"ur Festk\"orperphysik, Leibniz Universit\"at Hannover, Appelstra\ss e 2, 30167 Hannover, Germany}
\alsoaffiliation{Laboratorium f\"ur Nano- und Quantenengineering (LNQE), Leibniz Universit\"at Hannover, Schneiderberg 39, 30167 Hannover, Germany}
\email{mamiyev@fkp.uni-hannover.de}
\author{M. Tzschoppe}
\affiliation{Kirchhoff-Institut f\"ur Physik, Universit\"at Heidelberg, Im Neuenheimer Feld 227, 69120 Heidelberg, Germany}
\author{C. Huck}
\affiliation{Kirchhoff-Institut f\"ur Physik, Universit\"at Heidelberg, Im Neuenheimer Feld 227, 69120 Heidelberg, Germany}
\author{A. Pucci}
\affiliation{Kirchhoff-Institut f\"ur Physik, Universit\"at Heidelberg, Im Neuenheimer Feld 227, 69120 Heidelberg, Germany}
\author{H. Pfn\"ur} 
\affiliation{Institut f\"ur Festk\"orperphysik, Leibniz Universit\"at Hannover, Appelstra\ss e 2, 30167 Hannover, Germany}
\alsoaffiliation{Laboratorium f\"ur Nano- und Quantenengineering (LNQE), Leibniz Universit\"at Hannover, Schneiderberg 39, 30167 Hannover, Germany}
\email{pfnuer@fkp.uni-hannover.de}

\title{Plasmon Standing Waves by Oxidation of Si(553)-Au\footnotemark[1]}

\begin{document}

\begin{abstract}
Self-assembled Au atomic wires on stepped Si surfaces are metallic, as evidenced by a one-dimensionally dispersing plasmonic excitation.  Here we investigate the effects of oxidization on metallicity along such Au atomic wires on a regularly stepped Si(553) surface, by employing infrared absorption and high resolution electron energy loss spectroscopies. Our results indicate that only the Si environment undergoes oxidation, which has a remarkably small effect on the plasmon dispersion.  Only close to  $k_{\parallel}\rightarrow 0$ the plasmon dispersion ends at increasingly higher energies as a function of oxygen exposure, which is attributed to standing wave formation on small sections of Au wires generated by the introduction of O atoms as scattering centers, not to electronic gap opening.  This interpretation is in full agreement with the findings by infrared spectroscopy and with low energy electron diffraction.  
\end{abstract}
environment of metallic chains by adsorption and in particular by oxidation. 

\footnotetext[1]{This document is the unedited Author’s version of a Submitted Work that was subsequently accepted for
 publication in the Journal of Physical Chemistry C, copyright © American Chemical Society after peer review. To access the final edited
 and published work see doi: 10.1021/acs.jpcc.9b01372}

%\maketitle
% Introduction
\section{Introduction}
%%%%%%%%%%%%%%%
Metal-induced atomic wires are the ultimate limit of long-range ordered  quasi-one-dimensional (quasi-1D) electronic systems and may serve as the smallest possible interconnects in future electronic circuits. 
Confinement of conductive electrons within these systems gives rise to quasi-1D metallic bands with dispersion along the chains and brings up 1D electron channels with high electron mobility \cite{Crain2006}. 
In this respect, self-organized gold atomic wires on Si(hhk) surfaces are particularly interesting due to their well defined and long-range ordered surface structures. 
Depending on vicinal angle and direction with respect to the Si(111) surface, Au atoms arrange themselves in a single or in double atomic rows per terrace \cite{Crain2004}. 
Although the structural and electronic properties of such atomic wires have been addressed frequently \cite{Riikonen2008,Aulbach2013,Krawiec2016,Hafke2016,Hotzel2017a,
Lichtenstein2018,Sanna2018}, the interaction between these ultimate wires and the environment still remains rather unexplored. 
Recent studies on Si(hhk)-Au surfaces show that the gold-related bands are no longer free-electron-like, since they are strongly hybridized with the Si surface electronic states, particularly close to E$_{F}$ \cite{Song2015,Lichtenstein2018, Sanna2018,Mamiyev2018a}. 
Hence, the nature of coupling between the wires and substrate defines the electronic properties of metallic chains. 

In this context, the Si(553) surface with a terrace width of 1.47 nm, separated by double steps of Si, is one of the most addressed systems. Upon adsorption of 0.48 monolayers (ML) of Au, the so-called high coverage (HCW) phase is formed that contains one double gold strand per terrace, leaving the honeycomb-like step edge sites free of gold \cite{Crain2004,Song2015,Sanna2018}. 
Apart from a pairing reconstruction of the Au chains that give rise to a $\times{2}$ periodicity along the chains, partial spin ordering at low temperatures is coupled with the formation of a $\times 3$ periodicity along the step edges \cite{Aulbach2013}. 
While there is still discussion about the details of spin arrangement \cite{Snijders2012, Braun2018}, the interaction between these spins on adjacent terraces is so small that a spin liquid is formed \cite{Hafke2016}. 
Furthermore, when evaporating 0.19 ML of Au onto Si(553) surface, only every second terrace contains a double gold chain forming the low coverage (LCW) phase with a wire separation of 3.27\, nm \cite{Song2015,Sanna2018}. 
On the Au-free alternate terraces two extra dangling bonds per unit cell remain compared to the HCW phase, which increase the surface chemical potential and make this surface more reactive, therefore. 
The double Au strands result in the formation of  a doublet of electronic bands for both phases that are Rashba split \cite{Erwin2010,Song2015}. However, only one of them seems to cross the Fermi level for both phases \cite{Song2015, Sanna2018}, giving rise to a single plasmon excitation. 
As obvious from these remarks, this system provides an excellent test ground for modifications of the 

As shown in several studies in the recent past on two-dimensional (2D) \cite{Nagao2008,Rugeramigabo2008} as well as on quasi-one-dimensional systems \cite{Nagao2006,Rugeramigabo2010,Krieg2015}, the formation of low-energy plasmons, which in the long wavelength limit go to zero excitation energy,  is a clear indicator of metallicity. This metallicity depends on the environment and can be altered or even tuned by the embedding material or by adsorption. In that sense, we extend here  previous studies of oxidation on Si(557)-Au \cite{Mamiyev2018} and hydrogenation of Si(553)-Au \cite{Mamiyev2018a}. 
These low-dimensional plasmons should not be mixed-up with edge plasmons of a 3D solid, usually called surface plasmons or surface plasmon-polaritons \cite{Rocca1995,Pitarke2007,Politano2015}, which have a finite excitation energy in the long wavelength limit.

Furthermore, we want to directly compare results obtained by  electron loss spectroscopy (HREELS) with those from infrared absorption, which is known to be sensitive only to infrared active modes in the limit $k_\parallel \rightarrow 0$ \cite{Nagao2010}. Because of finite wire lengths, the plasmon resonance becomes optically detectable due to the dipolar character of excitations \cite{Hotzel2017a,Michael2019}. 
It is thus a very valuable tool for complementing our HREELS study that is also important to discriminate between finite size effects of wire lengths and possible adsorbate induced band gap opening.

\section{Experimental Setup}
The experiments were carried out in two independent UHV chambers, both  at a base pressure of  $5\times10^{-11}$ mbar. One was equipped with an IR spectrometer and RHEED \cite{Hotzel2017a}. 
The other chamber was equipped with a high-resolution spot profile analysis low energy diffraction (SPA-LEED) system and an electron energy loss spectrometer that combines both high momentum resolution with high energy resolution (EELS-LEED) \cite{Claus1992}. 
The EELS-LEED yields 0.01 {\AA}$^{-1}$ momentum and 12 meV energy resolution. 
Pieces of a  Si(553) wafer oriented 12.5$^{\circ}$ away from the (111) surface towards the [11$\overline{2}$] direction with a precision of $\pm 0.3^\circ$ were used for the investigations ($\rho_{Si(553)}\approx{0.01}\, \Omega$cm, n-doped). 
For IR measurements low-doped samples ($\rho = 5 - 6\,\Omega$cm, p-doped) were used and both sides of the sample were polished.
 
After ex-situ cleaning it was degassed at 600 $^{\circ}$C overnight, followed by flash-annealing up to 1250 $^{\circ}$C by direct current heating, while maintaining a pressure lower than 2${\times}$10$^{-10}$\,mbar. 
The successful preparation and the comparability was proven by SPA-LEED (EELS) and RHEED (IR) measurements. 

\begin{figure*}[tb]
\includegraphics[width=0.9\textwidth]{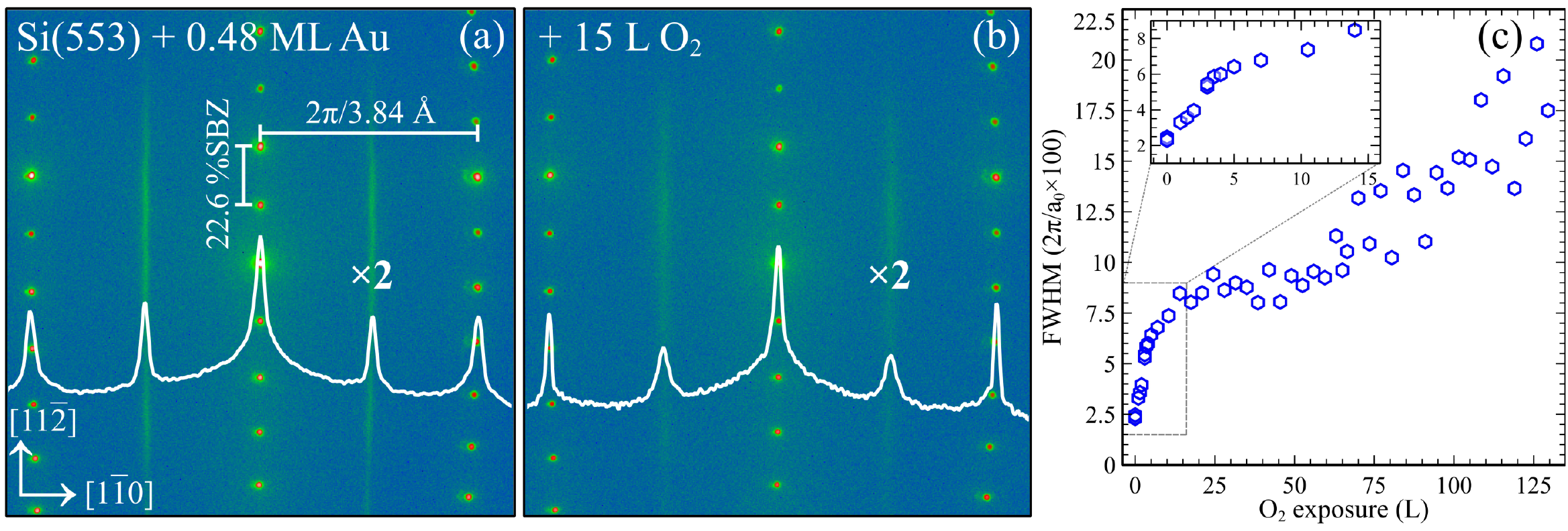} 
\caption{SPA-LEED patterns of the HCW phase of  Si(553)-Au (a) and after oxidation (b)  
Primary energy: 138 eV. White lines: Line scans along $[1\bar{1}0]$ normalized to the central intensity. c) Dependence of the full widths at half maximum of the $\times{2}$ streaks along the $[1\bar{1}0]$ direction on oxygen exposure.  Inset: magnification of the initial section. 
\label{Fig1}}
\end{figure*}

In the EELS chamber, nanowires were prepared by evaporating 0.19 ML (LCW) and 0.48 ML (HCW) of Au onto the flash-annealed surface at 630 $^{\circ}$C. After preparation, the sample was annealed to 930 $^{\circ}$C for $<1$ s to increase the atomic ordering. The last step produces a regular array of  (111) terraces with a separation of 14.7\,{\AA} (for the HCW phase)  (see the Fig\,\ref{Fig1}), separated by double-steps oriented along the $[1\bar{1}0]$-direction, consistent with previous works \cite{Song2015,Mamiyev2018a}. This annealing step was skipped for the LCW phase.

In the IR chamber, the LCW wires were prepared by evaporating 0.19 ML at 600 $^{\circ}$C onto the surface, followed by a post-annealing step at 870 $^{\circ}$C for 1 s. For the HCW phase, 0.48 ML were evaporated on the surface, which was kept at room temperature. A consecutive two-step post-annealing procedure (430 $^{\circ}$C for 10 s, 610 $^{\circ}$C for 5 s) finally led to the surface reconstruction. The Au deposition rate was kept at 0.1 ML/min for all surface preparation procedures.

The amount of Au was controlled by quartz microbalances. The density is given with respect to the Si layer density of the topmost plane (1 ML= 7.83$\times$10$^{14}$ atoms/cm$^{2}$) \cite{Claus1992,Sauerbrey1959}. 
Moreover, a SPA-LEED (or RHEED) pattern was taken after each preparation to monitor the sample quality. 
For adsorption measurements the chamber was back-filled with oxygen of $\approx{2\times{10}^{-8}}$\,mbar by a special gas line. 
Gas purity was checked via quadrupole mass spectrometers (QMS) in order to avoid any contamination with other gases. 
After exposure the oxygen flow was stopped and the pressure dropped immediately below ${2\times{10}^{-10}}$\,mbar. 
Only below this pressure IR and ELS-LEED measurements were started. 
This procedure helps to avoid an extra shift of spectra as a consequence of residual oxygen adsorption. The gas exposure is given in Langmuir (1L = 1$\times$10$^{-6}$\,mbar$\cdot$s).  

All measurements were carried out at room temperature. For EELS, an integration time of around $\approx{10}$ min per scan was chosen. The IR measurements were carried out by using a Fourier transform IR spectrometer (Tensor 27, Bruker) which is directly coupled to the UHV chamber by KBr windows. For acquiring spectra, a mercury cadmium telluride detector was used. All measurements (flash-annealed Si reference and Au nanowires) were scanned 200 times (measurement time: 86 s) with a resolution of 4 cm$^{-1}$ in transmittance geometry under normal incidence of light with either perpendicular or parallel polarized IR radiation (with respect to the Au wires). By dividing the spectra with wires by the reference spectra, the relative transmittance was obtained.
\section{Results and Discussion}

\subsection{Structural investigations}
The room temperature SPA-LEED pattern from the freshly prepared HCW phase is given in Fig.\ref{Fig1}a). 
The sharp integer order diffraction spots from Si(1x1) unit cell (attributed to step train) are clearly visible in both cases. 
The sharpness of these spots are indication of well arranged surfaces. $\times{2}$ streaks reflect dimerization of the Au double-chains along the [1$\overline{1}$0] direction.
 The spot splitting of 22.6\% SBZ (surface Brillouin zone) for the HCW phase corresponds to a periodicity of $4\frac{1}{3}a_\perp / \cos{12.5}^{\circ}$ ($a_\perp$=3.32 {\AA} is the substrate unit cell along the $[11\overline{2}]$-direction). 
 
 Interestingly, no qualitative changes are observed by adsorption of oxygen, even at an exposure of 200\,L (see, e.g., Fig.\,\ref{Fig1}b). Only the intensity of the $\times{2}$ streaks is reduced and their FWHM in $[1\bar{1}0]$ direction increases, as shown in more detail in Fig.\,\ref{Fig1}c) .
 In other words, oxygen adsorption distorts the long range order of the Au chains and breaks them up into smaller section. The reduction of the average chain lengths can be estimated from the inverse halfwidth (see below). 
 
 \begin{figure*}[tb]
\includegraphics[width=16cm]{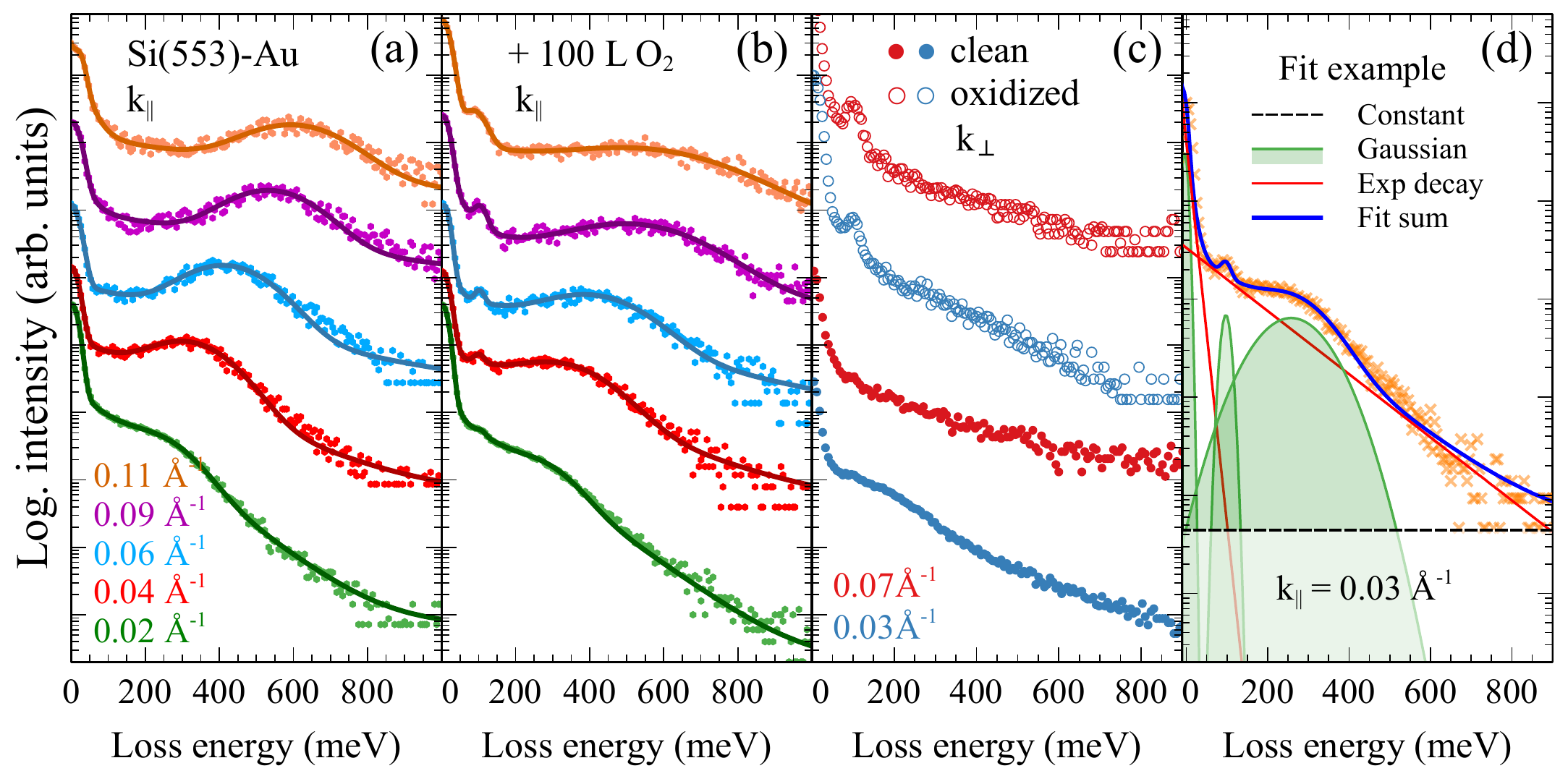} 
\centering
\caption{Electron energy loss sample spectra  from a) the clean Si(553)-Au HCW phase b) same after oxidation with 100 L of oxygen.  Both sets were measured at the $k_\parallel$  along the $[1\bar{1}0]$ direction, values indicated at the bottom left. c) Comparison on clean (bottom) and oxidized (top) EEL spectra at the two $k_\bot$ values indicated. d) Illustration of the fit routine (solid lines in a and b) applied to extract the plasmon dispersion. \label{Fig2}}
\end{figure*}

The LCW phase, in which only every second terrace is covered with a double strand of Au \cite{Song2015}, behaves  qualitatively quite similar, but with a higher sensitivity to oxygen. The large wire separation 32.7\,{\AA} minimizes the coupling between adjacent wires and results in the electronic structure described in refs. \cite{Song2015,Sanna2018}.
%%%%%%%%%%

\subsection{Plasmonic excitations and their modifications}
\begin{figure}[tb]
\includegraphics[width=7.5cm]{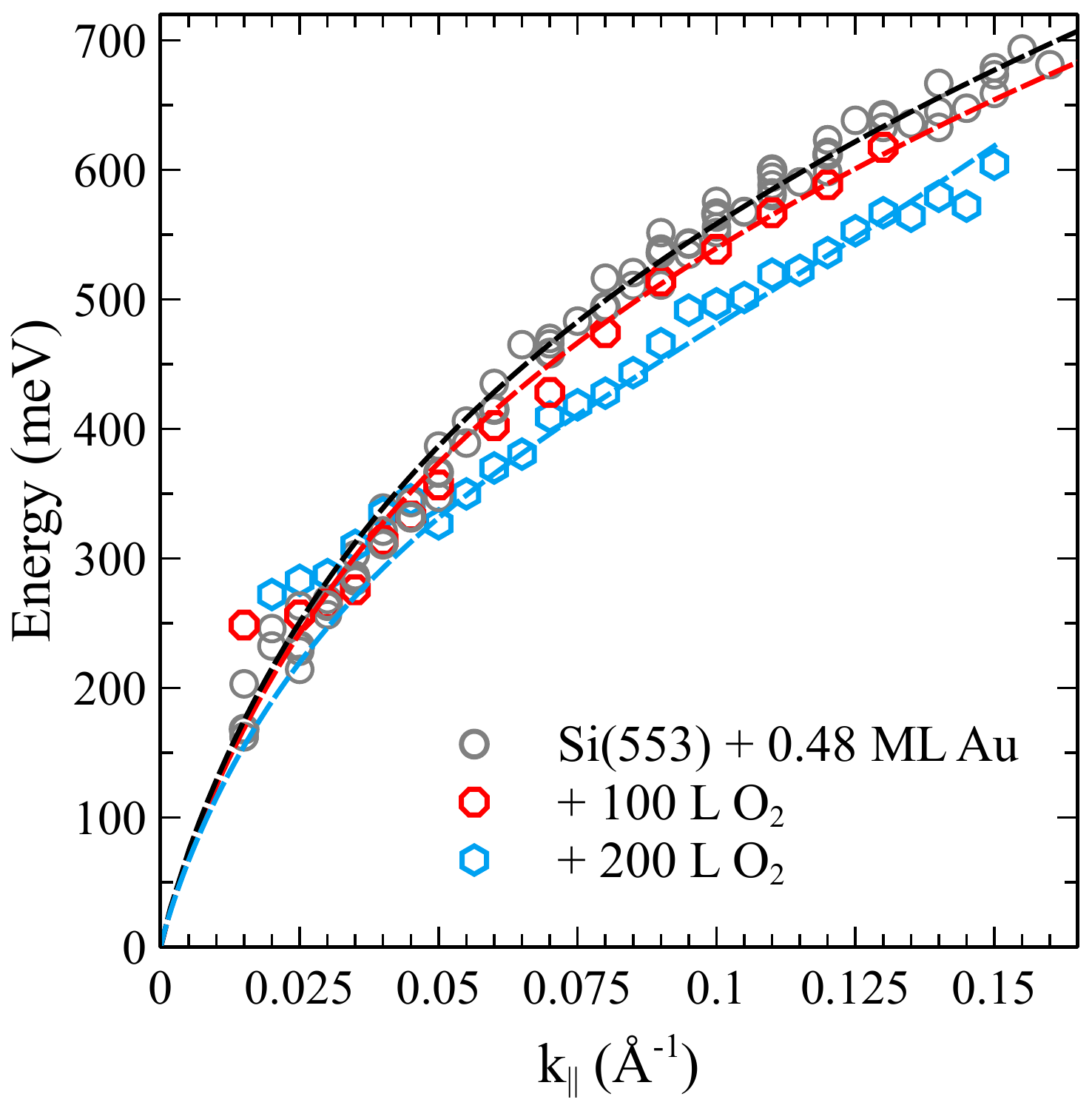} 
\centering
\caption{Plasmon dispersions for clean Si(553)-Au HCW phase and after various exposures to oxygen, as indicated. The data set  from ref.\,\cite{Sanna2018,Lichtenstein2016} was added in order to demonstrate the level of reproducibility. Solid black, red and blue lines: fits with the semi-empirical model according to ref.\,\cite{Lichtenstein2016} \label{Fig3}.}
\end{figure}
The typical loss spectra as a function of $k_\parallel$ from the HCW Si(553)-Au are given in Fig. \ref{Fig2}a). 
In both LCW and HCW phases this dispersing loss can only be observed along the Au wires.  
The exponential decay of background intensity as a function of energy  (see Fig.\,\ref{Fig2}) is known as the Drude tail  and is characteristic for the metallicity of these systems.  
An overview of the plasmonic excitations on the clean Si(553)-Au surface, demonstrating their close relation to the unoccupied bandstructure, has been given in ref.\, \cite{Sanna2018}. 

Upon oxidization a dispersionless loss peak at $\approx 100$\, meV emerges, which is identified with the Si-O stretch vibration normal to the surface \cite{Mamiyev2018}. It corresponds to stretch frequencies representing the initial stages of Si oxidation on their flat counterparts \cite{Ibach1982,Klevenz2010,niu2013}. 
The existence of this loss proves the presence of oxygen atoms on the surface, which also gets more intense with higher O$_{2}$ dosage. This peak is also clearly seen with infrared spectroscopy (see below). Its vibrational energy depends slightly on oxygen coverage, as discussed in ref. \cite{Klevenz2010}. Because IR transmittance at normal incidence of light is sensitive only to vibrational dipoles parallel to the surface, the Si-O bonds should have a oblique orientation on the surface.
The apparent increase in its intensity with increasing $k_{\parallel}$, however, is not real, since it is due to the normalization of the spectra to the elastic peak at a given $k_\parallel$.  

Previous studies with DFT of the oxidation of Si(557)-Au \cite{Edler2017} show that dissociation of oxygen molecules is not possible at Au sites. Even direct binding of oxygen atoms with the Au chains directly is very unfavorable, so that only the Si surface atoms will be oxidized.  Since the local structure of Si(553), apart from the Si adatom chain, is quite similar, this result should also hold qualitatively for the Si(553)-Au system. Thus, electronic changes of the wire properties should only occur by indirect interaction. 
%%%%%%%%%%%%%%%%%%%%%%%
\begin{figure}[tb]
\includegraphics[width=6.5cm]{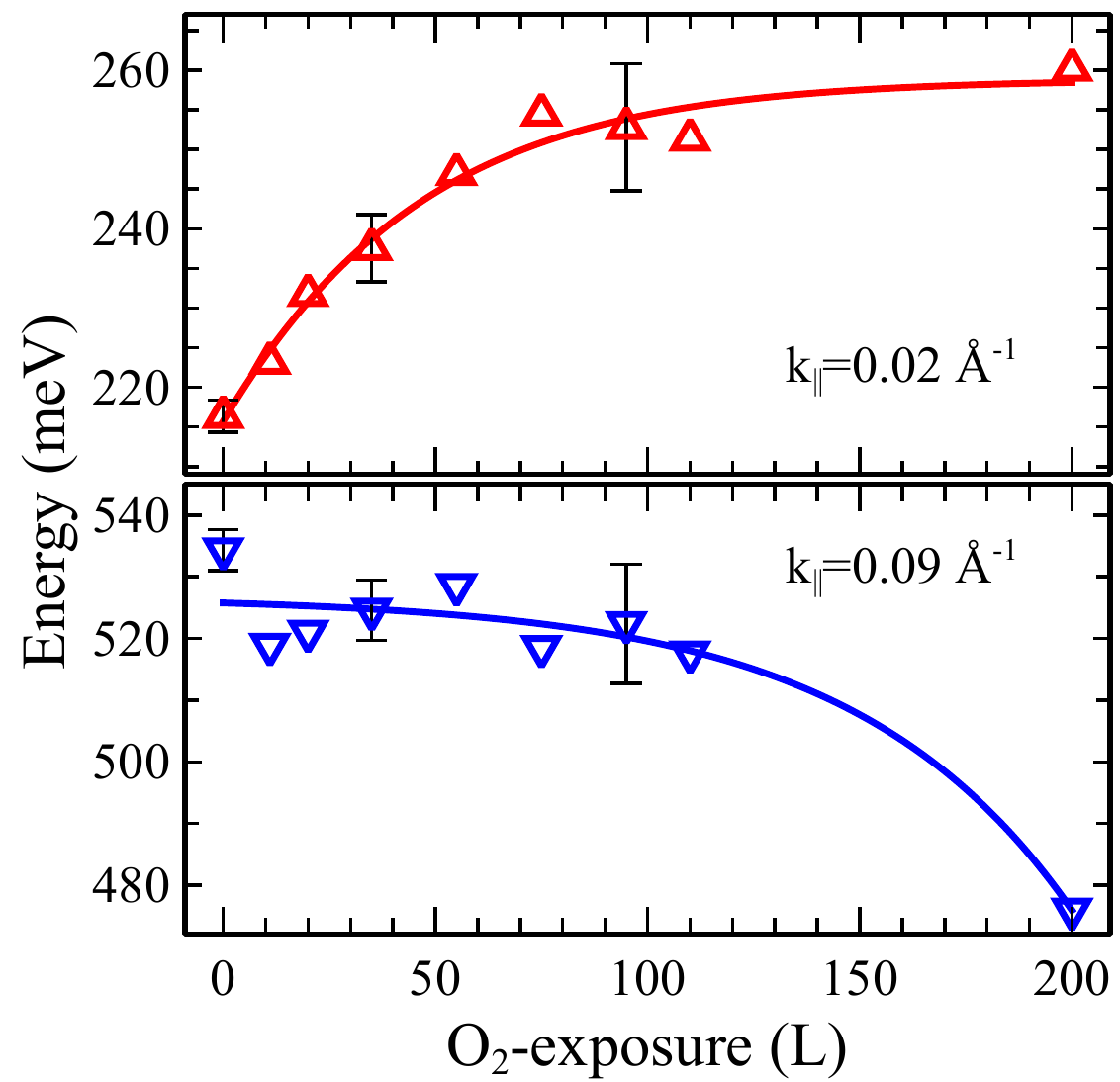} 
\caption{Changes of plasmon energy at defined $k_{\parallel}$ as a function of O$_{2}$ exposure on the HCW phase. \label{Fig4}}
\end{figure}
The plasmon dispersions for the clean system and after oxidization with various amounts of oxygen are shown in Fig.\ref{Fig3}. The solid lines starting at zero energy at $k_\parallel = 0$ are  fits  with a modified confined free electron gas model \cite{Lichtenstein2016}. 
However upon oxidization the plasmon dispersion deviates from this extrapolation at $k_{\parallel}{\rightarrow}0$ and the onset of deviation shifts to higher $k_{\parallel}$ with increasing O$_{2}$ dosage.
For $k_\parallel > 0.05$\,\AA$^{-1}$ dispersion is insensitive to oxidation up to an exposure of 100\,L within error margins characterized by the scatter of data points in Fig.\,\ref{Fig3}.  A general reduction of plasmon dispersion was only observed above an exposure of 100\,L, but even at an exposure to 200\,L was only around 15\% with respect to the clean surface.   

Comparing these results with a recent study on the oxidation of Si(557)-Au, there is significant similarity. In both cases metallicity withstands adsorption of a very significant amount of oxygen. 
At first glance, the Si(553)-Au system seems to be less reactive than Si(557)-Au, since much higher doses of oxygen are needed to induce any change of dispersion in the former system.  
However, this statement should be taken with caution, since in simulations of Si(557)-Au + O  system the plasmon dispersion turned out to be quite insensitive to oxidation of the Si honeycomb chain, which is the common structural element on both surfaces. \cite{Mamiyev2018}
In fact, the largest  changes in plasmon dispersion induced by oxygen adsorption were due to oxidation of the Si adatom chain, which does not exist in Si(553)-Au. Therefore, these results are fully compatible with the assumption that mainly the Si step edge with its local honeycomb structure (HC) is oxidized on Si(553)-Au, which then has a small and only indirect effect on the conductive properties of the Au chains, but does not mean that the reactivity with oxygen is low. 

This statement is corroborated by theoretical results of complete oxidation of the HC chain on Si(553)-Au \cite{Edler2017}, which even agrees semi-quantitatively with the slight reduction of the plasmon frequency measured at the oxygen exposure of 200 L.  
In the simulation, the  dispersion of the Au-induced unoccupied band is slightly lowered compared with the unoxidized surface and band gap opening at  500 meV above E$_F$ is seen, together with possibly a small band gap opening ($<100$\,meV) close to E$_F$. 
This opening of a small band gap may be the reason for the reduction of measurable DC conductivity at essentially the isotropic background level for oxygen exposures around 200 L O$_{2}$ \cite{Edler2017}. 

If we extrapolate our measured plasmon dispersion to $k_\parallel = 0$ at the highest oxygen exposure, we end up at a value of 280 meV, i.e. at a much higher value than that expected from such a small band gap.  
In other words, this high extrapolated value is not dominated by a band gap, but is more likely due to disorder induced by oxygen adsorption. 

Inevitably oxygen adsorption, which occurs in a random fashion, as concluded from our LEED results, induces disorder in the system. 
The consequence is an increased scattering rate also for plasmons, which on the wires can only be backscattered. 
Therefore, standing plasmonic waves on shorter and shorter sections of the Au wires are formed by increasing concentrations of oxygen, and dispersing plasmons can only exist for shorter  wavelengths than these limits. As a result, the dispersion close to $k_\parallel = 0$ disappears for  larger and larger sections. This behavior is reflected by an increase of plasmon loss energy as a function of oxygen exposure at fixed small $k_\parallel$, as shown in the top panel of  Fig.\ref{Fig4} for $k_\parallel = 0.02 $\,\AA. This contrasts with the behavior at large k, where the plasmon loss was found to be constant over a wide range of exposures and finally decreases  (see Fig. \,\ref{Fig4} at $k_{\parallel}=0.09 ${\AA}$^{-1}$) above 100\,L of O$_{2}$.

\subsection{Plasmon resonance: oxygen-induced standing wave formation}
\begin{figure}[tb]
\includegraphics[width=8cm]{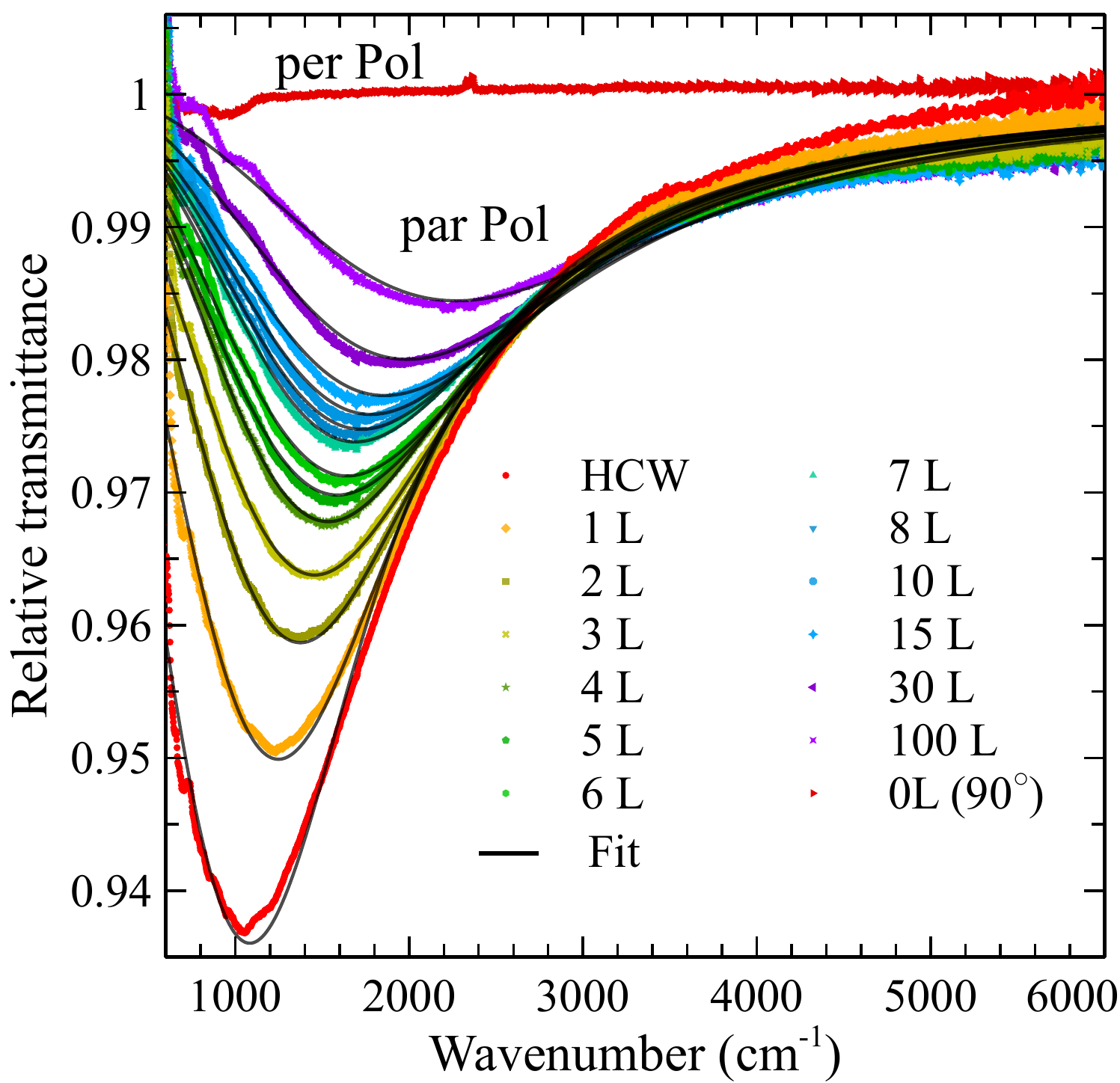} 
\centering
\caption{Relative transmittance spectra of clean and oxidized Au-Si(553) surfaces for the HCW phase. 
Relative spectra were obtained by referencing the spectrum to clean Si(553) surface (measured before Au evaporation) from the spectra of Au wires on Si. Solid lines represent the fit from the formula as described in text. \label{Fig5}}
\end{figure}
The interpretation of the EELS data at small $k_\parallel$ is supported by infrared transmittance experiments.

As a function of oxygen exposure we see a gradual shift of the absorption minimum to higher wavenumbers, accompanied by successive broadening. 
As already suggested for the pure system \cite{Hotzel2017a,Neubrech2006}, the absorption minimum is mainly sensitive to the average chain length that is relevant for plasmon propagation, while  the absorption width also reflects changes in the length distribution. Therefore, the observed shift of the absorption minimum indicates successive shortening of the average wire length, while the increasing absorption width is indicative for the widening of the distribution of wire length as a function of O$_2$ exposure. 
The resulting relative transmittance IR spectra from the as prepared and oxygen exposed Si(553)-Au surfaces  are presented in Figs.\, \ref{Fig5} (HCW) and \ref{Fig8} (LCW), respectively.  Strong absorption is only seen for polarization parallel to the wires (in the [1$\overline{1}0$] direction), which reflects the anisotropic metallicity of self-assembled Au wires. The observed absorption feature is associated with the low-energy plasmonic excitation in metal-induced atomic chains, as studied previously \cite{Nagao2006,Hotzel2017a,Lichtenstein2018, Sanna2018} for different surfaces.  The weak absorption signal around 950 cm$^{-1}$ appears upon oxygen adsorption. 
It  is attributed to Si-O stretch mode and provides evidence for  dissociative adsorption of oxygen \cite{Ibach1982,Klevenz2010}.
We first concentrate on the gradual shift of IR absorption minima and interpret this shift as being due to an effective shortening of wires by oxidation. 
The spectral shape, especially the plasmonic resonance position of the HCW wires was investigated by fitting the data according to ref.\,\cite{Michael2019}. 
With this assumption we identify the energy of the IR absorption minima with the energy of the ground state of a plasmon standing wave on wires with an average length ${l}$. 
This wavelength cannot be exceeded at a given oxygen concentration so that no dispersion is possible for smaller $k_\parallel$. 
Therefore, we took the energies of the IR absorption minima  and determined the corresponding $k_\parallel$ value as the intersection of a non-dispersing state of this energy with the actual dispersion shown in Fig.\,\ref{Fig3}. We further assume that the wavelength at this k-value is that of the plasmonic ground state, i.e. the wavelength is twice as long as the average wire.  

\begin{figure}[tb]
\includegraphics[width=1\columnwidth]{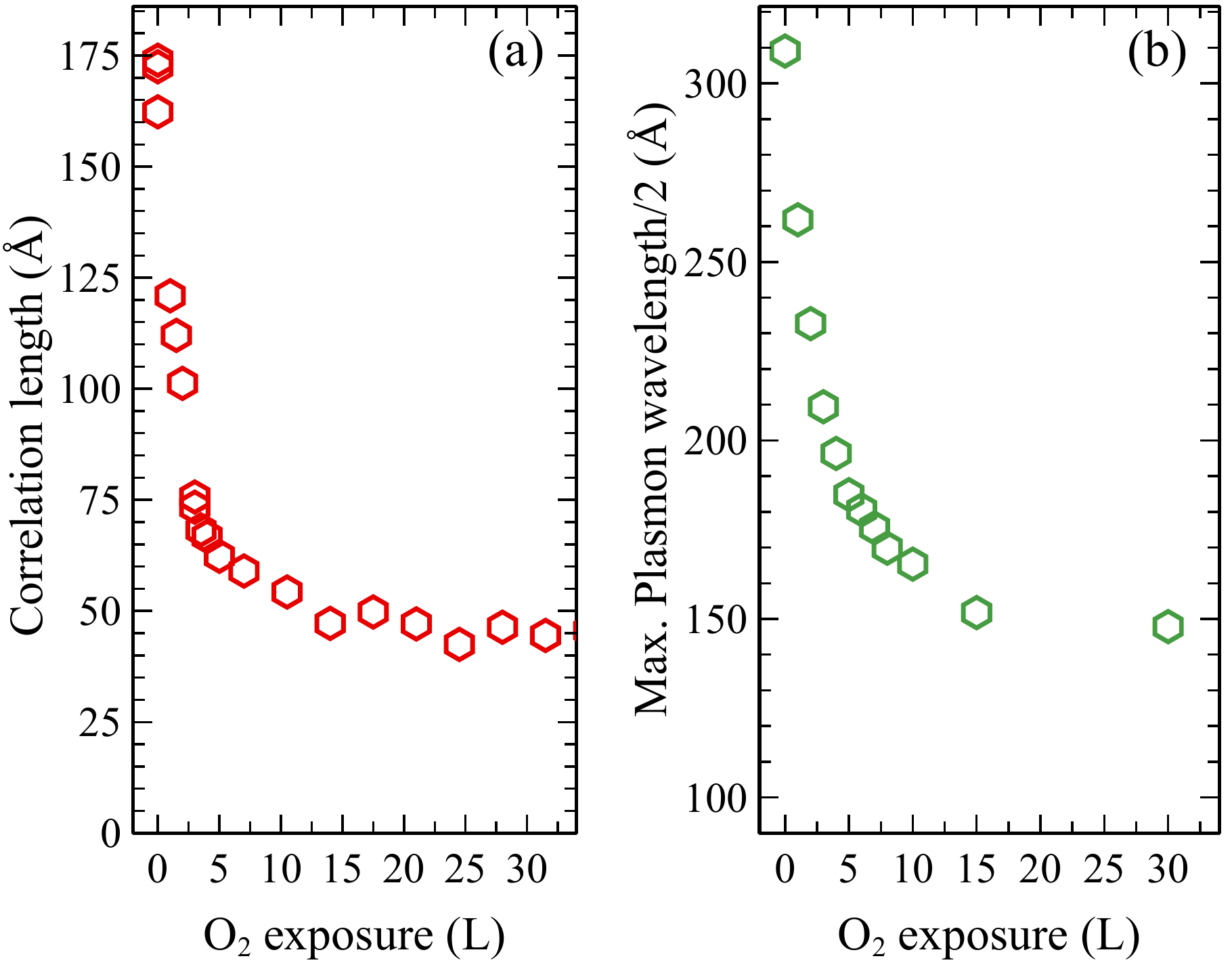} 
%\centering
\caption{a) Correlation lengths of $\times{2}$ streaks as a function of oxygen exposure in Si(553)-Au, determined as the inverse FWHMs from the  line scans shown in Fig.\,\ref{Fig1}. b) Maximum plasmon wavelengths divided by 2, determined from the energetic positions of the absorption minima (Fig.\,\ref{Fig5} in conjunction with the plasmon dispersion (Fig.\,\ref{Fig3}). For details, see text.  
\label{Fig6}
}
\end{figure}

This average wire length is plotted in Fig.\,\ref{Fig6}b) as half of this maximum wavelength.
The initial value of about 31 nm effective length without oxygen is in good agreement with the result obtained in ref.\,\cite{Hotzel2017a}. 
Oxidation, which happens mainly on the HC Si chain \cite{Edler2017}, still has a significant influence also on plasmon scattering by defects, which strongly increases so that the effective wire length for plasmons decreases correspondingly, as shown in this figure. 

A very similar trend is seen when taking the inverse halfwidths (FWHM) of the $\times 2$ streaks along the $[1\bar{1} 0]$ direction in LEED (see Fig.\,\ref{Fig6}a)) from measurements similar to those in Fig. \,\ref{Fig1}. 
The reduction of the correlation length of $\times 2$ order as a function of oxygen exposure is even more pronounced than in IR and starts at a significantly smaller value, indicating different sensitivity of IR and LEED to various types of disorder introduced by oxygen atoms. It is noteworthy to mention that the small change in band structure in this case is shifting the resonance to longer wavelength- and thus the used approximetion may deliver a little bit too long lengths. 

These different sensitivities to defects can at least be qualitatively understood. Plasmons generally have a wavelength, $\lambda \gg a$ (a is the lattice constant), and exhibit a pronounced wavelength dependence \cite{Langer2010} that makes the scattering probability at defects of atomic size small at long wavelengths. 
On the contrary, stacking faults in the dimerization of the Au chains, e.g., clearly cause phase changes in LEED and thus limit the correlation length in LEED, whereas they are expected to represent weak scatterers for the plasmons with wavelengths. 
Kinks in the Au chains, on the other hand, caused by roughness of step edges or slight azimuthal misalignment of the sample, should also scatter plasmons more efficiently. Therefore, we always expect a higher sensivitity to structural defects  in LEED than for plasmons.  

This scattering scenario remains essentially unchanged when the system is oxidized: As outlined above, oxidation happens mainly on the (insulating) Si HC chains, i.e. it only indirectly influences the conducting channel by possible local relaxations of the Si atoms involved in oxygen bonding. 
Thus  it influences the local effective wire width and/or causes distortions of the bandstructure close to E$_F$. 
Again, the geometric distortions by disordered oxygen adsorption seem to be more drastic compared with the electronic modifications, as concluded from a comparison of the FWHM in LEED with the  plasmonic dispersion curve, which turns  out to be virtually unchanged up to an exposure of about 100 \,L. Therefore, the random geometric disorder reflected in the reduced correlation length due to O  adsorption in LEED shows a similar tendency as the effective wire length for plasmons, but the plasmonic effective wire lengths are larger  by a factor of 2.5 to 3.  
The widening of the IR absorption spectrum, on the other hand, is an indication of electronic damping which reduces absorption and also broadens the resonance. 

\subsection{The LCW phase}
\begin{figure}[tb]
\includegraphics[width=7.25cm]{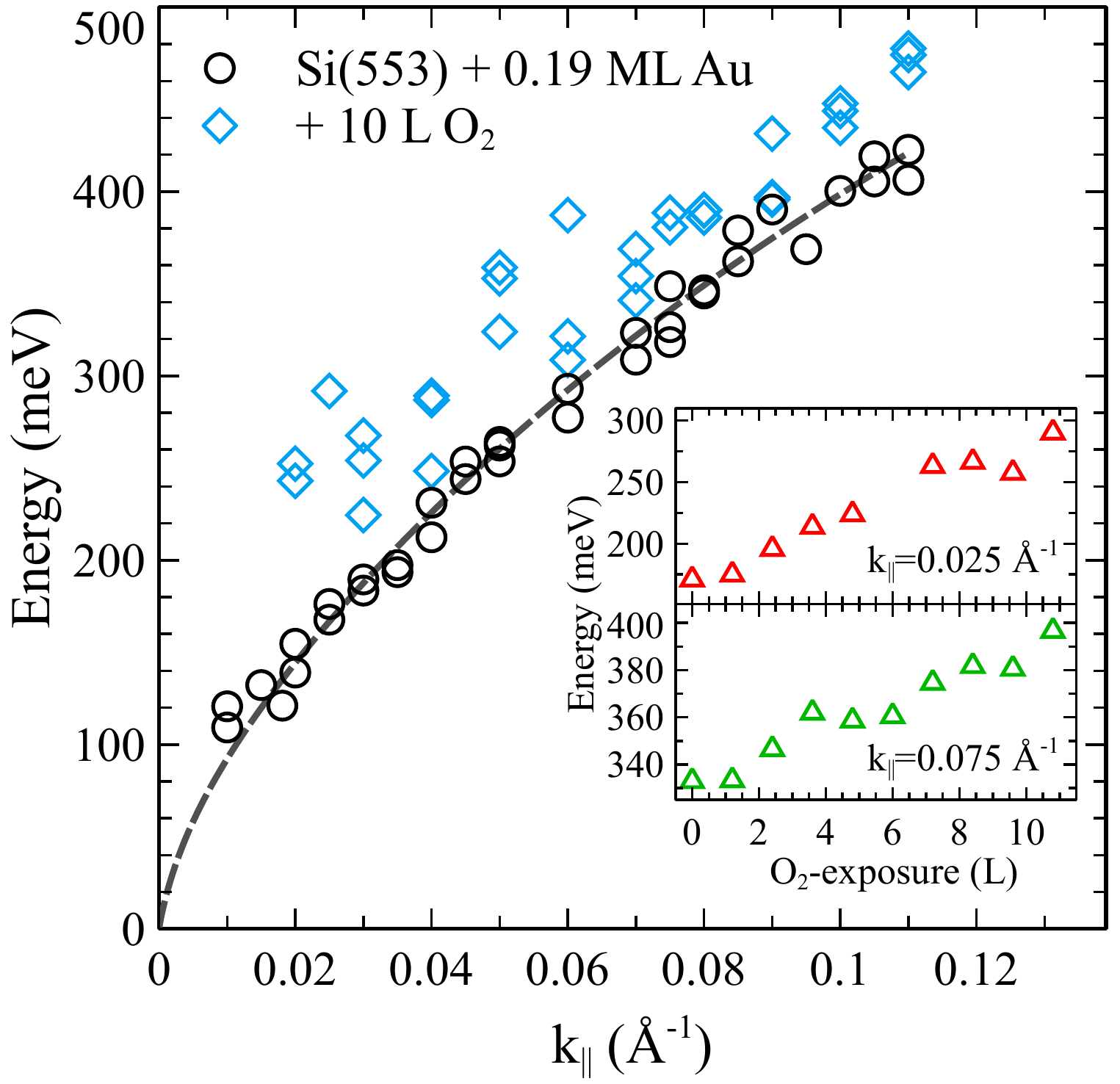} 
\centering
\caption{Plasmon dispersion of the clean and oxidized LCW phase obtained from the loss maxima in EELS.  The inset shows the evalution of plasmon frequency as a function of O$_{2}$ dosage at two fixed values of  $k_{\parallel}$. \label{Fig7}}
\end{figure}

The plasmons in the LCW phase turn out to be more sensitive to oxidation. 
The weaker plasmon signal of this phase compared with HCW (see Figs.\,\ref{Fig2} and \,\ref{Fig5}) can easily be understood, since only 4/9 of the surface consist of terraces hosting a gold wire. 
A further loss of intensity is caused by the lower quality of ordering that is obvious from the diffraction profiles (not shown). On the other hand, Si adatoms with unsaturated bonds at the uncovered Si(1$\times$1) terraces make this surface chemically more reactive than the HCW surface. 
For these reasons only 30\% of the initial loss intensity remains after an exposure of 10 \,L O$_{2}$, which made it barely detectable at higher exposures.
By applying the same fit routine as shown in Fig.\,\ref{Fig2}, the plasmon dispersion shown in Fig.\,\ref{Fig7}  was obtained. 

\begin{figure}[tb]
\includegraphics[width=8.5cm]{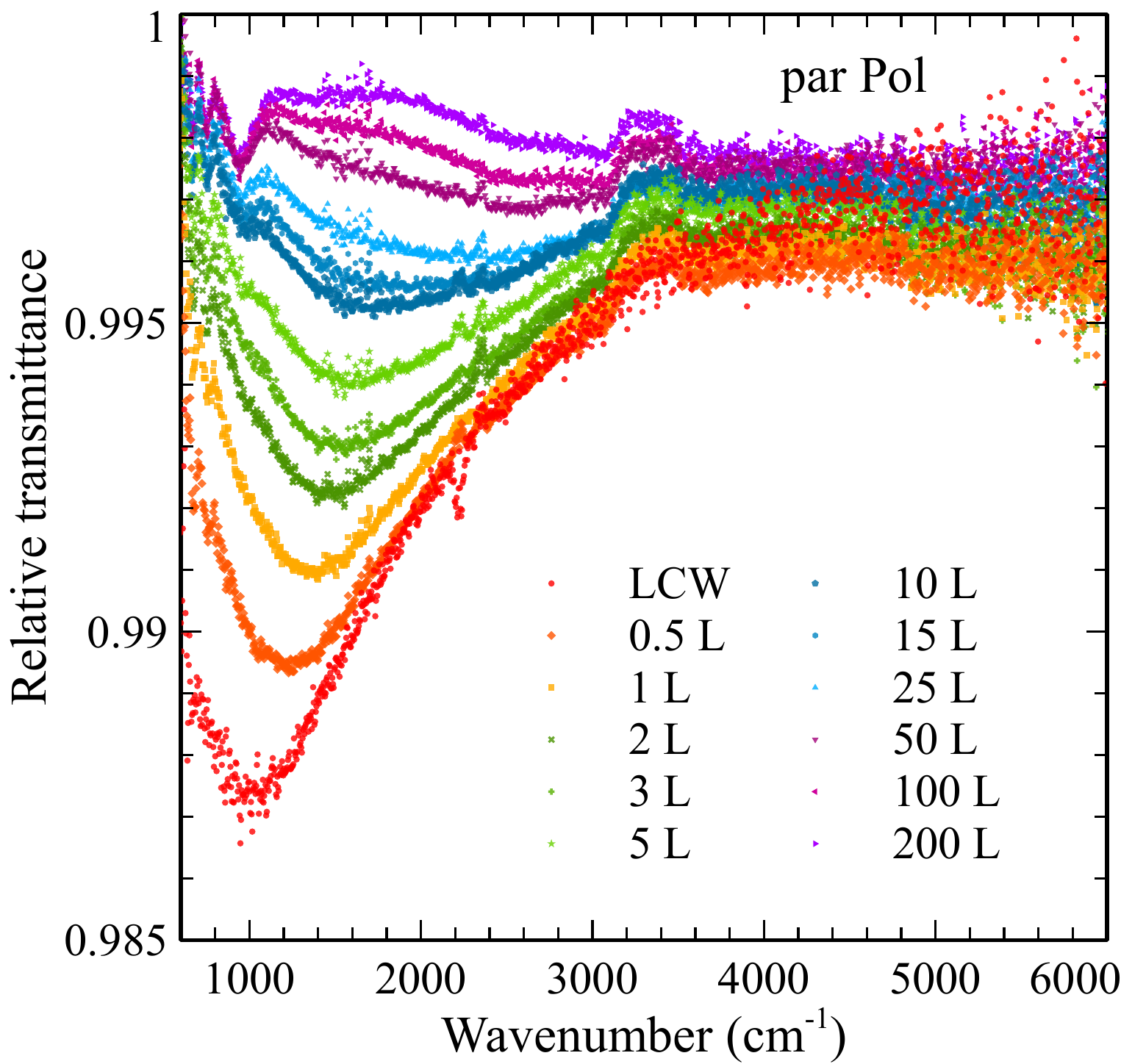} 
\centering
\caption{Relative transmittance spectra at normal incidence with polarization along the wires from the clean Si(553)-Au LCW phase and after oxygen exposure, as indicated.\label{Fig8}}
\end{figure}

At small $k_\parallel$ an increase of plasma frequency was seen as a function of oxygen exposure that is  very similar to that in  the HCW phase (see Figs.\,\ref{Fig5} and \ref{Fig8}). Therefore, it must have an analogous origin.  
Only the sensitivity to oxygen exposure is higher, which limits the possible range of exposures in EELS to 10 \,L.  This higher sensitivity is also reflected in oxygen-induced changes of the plasmon dispersion (Fig.\,\ref{Fig7}). 
Contrary to the HCW phase, however, O adsorption tends to increase the plasmon dispersion
(see inset in Fig.\,\ref{Fig7}).

While changes in band structure close to E$_F$ are expected to be very similiar in both phases, qualitative differences may be due to the following mechanisms:  
First, the reduction of the dielectric function should be more significant in the LCW phase, since a much larger fraction of the surface is pure Si, which is primarily oxidized to form SiO$_x$. Second, the relaxation of surface tension by oxidation in the uncovered terraces may induce additional confinement normal to wires.  

The qualitative behavior of IR absorption spectroscopy at normal incidence as a function of oxygen exposure is similar as for the HCW phase (Fig.\,\ref{Fig5}), and is in agreement with the interpretation of the small k-behavior given there. 
Also in IR the absorption is significantly smaller than for the HCW phase for the reasons already discussed above. Compared with  EELS, an absorption signal is detectable up to considerably higher oxygen exposures. Interestingly, any plasmonic absorption disappears for exposures above 50 L, contrary to the HCW phase, while the absorption due to the Si-O stretch is still clearly visible. Therefore, we conclude  that metallicity is in fact destroyed for oxygen exposures above 50 L. Whether this finding is due to extreme disorder or still a bandstructure effect has to remain an open question at this point. 

\section{Summary and conclusions}
Our investigations, using a combination of HREELS, IR spectroscopy and LEED, demonstrate valuable complementarity of methods. Specificly for the systems investigated here, we showed  that metallicity in the high coverage phase of Au on Si(553), containing a double wire of Au on each terraces, is quite robust, as judged from the small changes of plasmon dispersion upon oxidation. It is, however, susceptible to disorder introduced by the adsorption of oxygen mainly in the honeycomb chain of Si at the step edge, with leads to formation of standing plasmonic waves.  While opening of a small bandgap close to E$_F$ cannot be fully excluded, it does not dominate the measurable plasmon dispersion.  

The low coverage phase, however, is much more sensitive to oxidation. While similar behavior as in the HCW phase was found for small oxygen exposures, there is an overall increase of plasmon dispersion, which can be rationalized. The complete disappearance of the plasmonic absorption signal in IR at high oxygen exposures, however, demonstrates final destruction of metallicity of this phase.  

\begin{acknowledgement}
We gratefully acknowledge financial support from the Deutsche Forschungsgemeinschaft in the research unit FOR 1700 and Niedersächsisches Ministerium für Wissenschaft und Kultur through the graduate school ‘contacts in nanosystems’. Heidelberg: Funding by collaborative research center SFB 1246. M. T. acknowledges support from the Heidelberg Graduate School for Fundamental Physics (HGSFP).  
\end{acknowledgement}

%************************************************************************
%\bibliographystyle{abbrv}
%\bibliographystyle{achemso}
\bibliography{library-plasmons-19-01-18}

\providecommand{\latin}[1]{#1}
\makeatletter
\providecommand{\doi}
  {\begingroup\let\do\@makeother\dospecials
  \catcode`\{=1 \catcode`\}=2 \doi@aux}
\providecommand{\doi@aux}[1]{\endgroup\texttt{#1}}
\makeatother
\providecommand*\mcitethebibliography{\thebibliography}
\csname @ifundefined\endcsname{endmcitethebibliography}
  {\let\endmcitethebibliography\endthebibliography}{}
\begin{mcitethebibliography}{35}
\providecommand*\natexlab[1]{#1}
\providecommand*\mciteSetBstSublistMode[1]{}
\providecommand*\mciteSetBstMaxWidthForm[2]{}
\providecommand*\mciteBstWouldAddEndPuncttrue
  {\def\EndOfBibitem{\unskip.}}
\providecommand*\mciteBstWouldAddEndPunctfalse
  {\let\EndOfBibitem\relax}
\providecommand*\mciteSetBstMidEndSepPunct[3]{}
\providecommand*\mciteSetBstSublistLabelBeginEnd[3]{}
\providecommand*\EndOfBibitem{}
\mciteSetBstSublistMode{f}
\mciteSetBstMaxWidthForm{subitem}{(\alph{mcitesubitemcount})}
\mciteSetBstSublistLabelBeginEnd
  {\mcitemaxwidthsubitemform\space}
  {\relax}
  {\relax}

\bibitem[Crain(2006)]{Crain2006}
Crain,~J.~N. {Low-dimensional electronic states at silicon surfaces}.
  \emph{Appl. Phys. A} \textbf{2006}, \emph{82}, 431--438\relax
\mciteBstWouldAddEndPuncttrue
\mciteSetBstMidEndSepPunct{\mcitedefaultmidpunct}
{\mcitedefaultendpunct}{\mcitedefaultseppunct}\relax
\EndOfBibitem
\bibitem[Crain \latin{et~al.}(2004)Crain, McChesney, Zheng, Gallagher,
  Snijders, Bissen, Gundelach, Erwin, and Himpsel]{Crain2004}
Crain,~J.~N.; McChesney,~J.; Zheng,~F.; Gallagher,~M.~C.; Snijders,~P.;
  Bissen,~M.; Gundelach,~C.; Erwin,~S.~C.; Himpsel,~F.~J. {Chains of gold atoms
  with tailored electronic states}. \emph{Phys. Rev. B} \textbf{2004},
  \emph{69}, 125401\relax
\mciteBstWouldAddEndPuncttrue
\mciteSetBstMidEndSepPunct{\mcitedefaultmidpunct}
{\mcitedefaultendpunct}{\mcitedefaultseppunct}\relax
\EndOfBibitem
\bibitem[Riikonen and S{\'{a}}nchez-Portal(2008)Riikonen, and
  S{\'{a}}nchez-Portal]{Riikonen2008}
Riikonen,~S.; S{\'{a}}nchez-Portal,~D. {Systematic investigation of the
  structure of the Si(553)-Au surface from first principles}. \emph{Phys. Rev.
  B} \textbf{2008}, \emph{77}, 165418\relax
\mciteBstWouldAddEndPuncttrue
\mciteSetBstMidEndSepPunct{\mcitedefaultmidpunct}
{\mcitedefaultendpunct}{\mcitedefaultseppunct}\relax
\EndOfBibitem
\bibitem[Aulbach \latin{et~al.}(2013)Aulbach, Sch{\"{a}}fer, Erwin, Meyer,
  Loho, Settelein, and Claessen]{Aulbach2013}
Aulbach,~J.; Sch{\"{a}}fer,~J.; Erwin,~S.~C.; Meyer,~S.; Loho,~C.;
  Settelein,~J.; Claessen,~R. {Evidence for long-range spin order instead of a
  Peierls transition in Si(553)-Au chains}. \emph{Phys. Rev. Lett.}
  \textbf{2013}, \emph{111}, 137203\relax
\mciteBstWouldAddEndPuncttrue
\mciteSetBstMidEndSepPunct{\mcitedefaultmidpunct}
{\mcitedefaultendpunct}{\mcitedefaultseppunct}\relax
\EndOfBibitem
\bibitem[Krawiec \latin{et~al.}(2016)Krawiec, Kopciuszy{\'{n}}ski, and
  Zdyb]{Krawiec2016}
Krawiec,~M.; Kopciuszy{\'{n}}ski,~M.; Zdyb,~R. {Different spin textures in
  one-dimensional electronic bands on Si(5 5 3)-Au surface}. \emph{Appl. Surf.
  Sci.} \textbf{2016}, \emph{373}, 26--31\relax
\mciteBstWouldAddEndPuncttrue
\mciteSetBstMidEndSepPunct{\mcitedefaultmidpunct}
{\mcitedefaultendpunct}{\mcitedefaultseppunct}\relax
\EndOfBibitem
\bibitem[Hafke \latin{et~al.}(2016)Hafke, Frigge, Witte, Krenzer, Aulbach,
  Sch{\"{a}}fer, Claessen, Erwin, and {Horn-von Hoegen}]{Hafke2016}
Hafke,~B.; Frigge,~T.; Witte,~T.; Krenzer,~B.; Aulbach,~J.; Sch{\"{a}}fer,~J.;
  Claessen,~R.; Erwin,~S.~C.; {Horn-von Hoegen},~M. {Two-dimensional
  interaction of spin chains in the Si(553)-Au nanowire system}. \emph{Phys.
  Rev. B} \textbf{2016}, \emph{94}, 161403\relax
\mciteBstWouldAddEndPuncttrue
\mciteSetBstMidEndSepPunct{\mcitedefaultmidpunct}
{\mcitedefaultendpunct}{\mcitedefaultseppunct}\relax
\EndOfBibitem
\bibitem[H{\"{o}}tzel \latin{et~al.}(2017)H{\"{o}}tzel, Galden, Baur, and
  Pucci]{Hotzel2017a}
H{\"{o}}tzel,~F.; Galden,~N.; Baur,~S.; Pucci,~A. {One-dimensional plasmonic
  excitations in gold-induced superstructures on Si(553): Impact of gold
  coverage and silicon step edge polarization}. \emph{J. Phys. Chem. C}
  \textbf{2017}, \emph{121}, 8120\relax
\mciteBstWouldAddEndPuncttrue
\mciteSetBstMidEndSepPunct{\mcitedefaultmidpunct}
{\mcitedefaultendpunct}{\mcitedefaultseppunct}\relax
\EndOfBibitem
\bibitem[Lichtenstein \latin{et~al.}(2018)Lichtenstein, Mamiyev, Sanna,
  Schmidt, Tegenkamp, and Pfn{\"{u}}r]{Lichtenstein2018}
Lichtenstein,~T.; Mamiyev,~Z.; Sanna,~S.; Schmidt,~W.~G.; Tegenkamp,~C.;
  Pfn{\"{u}}r,~H. {Probing quasi-1D band structures by plasmon spectroscopy}.
  \emph{Phys. Rev. B} \textbf{2018}, \emph{97}, 165421\relax
\mciteBstWouldAddEndPuncttrue
\mciteSetBstMidEndSepPunct{\mcitedefaultmidpunct}
{\mcitedefaultendpunct}{\mcitedefaultseppunct}\relax
\EndOfBibitem
\bibitem[Sanna \latin{et~al.}(2018)Sanna, Lichtenstein, Mamiyev, Tegenkamp, and
  Pfn\"ur]{Sanna2018}
Sanna,~S.; Lichtenstein,~T.; Mamiyev,~Z.; Tegenkamp,~C.; Pfn\"ur,~H. How
  one-dimensional are atomic Au chains on a substrate? \emph{J. Phys. Chem. C}
  \textbf{2018}, \emph{122}, 25580--25588\relax
\mciteBstWouldAddEndPuncttrue
\mciteSetBstMidEndSepPunct{\mcitedefaultmidpunct}
{\mcitedefaultendpunct}{\mcitedefaultseppunct}\relax
\EndOfBibitem
\bibitem[Song \latin{et~al.}(2015)Song, Oh, Shin, Ahn, Moon, Woo, Choi, Park,
  and Ahn]{Song2015}
Song,~I.; Oh,~D.-H.; Shin,~H.-C.; Ahn,~S.-J.; Moon,~Y.; Woo,~S.-H.;
  Choi,~H.~J.; Park,~C.-Y.; Ahn,~J.~R. {Direct momentum-resolved observation of
  one-dimensional confinement of externally doped electrons within a single
  subnanometer-scale wire}. \emph{Nano Lett.} \textbf{2015}, \emph{15},
  281--288\relax
\mciteBstWouldAddEndPuncttrue
\mciteSetBstMidEndSepPunct{\mcitedefaultmidpunct}
{\mcitedefaultendpunct}{\mcitedefaultseppunct}\relax
\EndOfBibitem
\bibitem[Mamiyev \latin{et~al.}(2018)Mamiyev, Sanna, Lichtenstein, and
  Pfn\"ur]{Mamiyev2018a}
Mamiyev,~Z.; Sanna,~S.; Lichtenstein,~C.,~T. and.~Tegenkamp; Pfn\"ur,~H.
  {Extrinsic doping on the atomic scale: tuning metallicity in atomic Au
  chains}. \emph{Phys. Rev. B} \textbf{2018}, \emph{98}, 245414\relax
\mciteBstWouldAddEndPuncttrue
\mciteSetBstMidEndSepPunct{\mcitedefaultmidpunct}
{\mcitedefaultendpunct}{\mcitedefaultseppunct}\relax
\EndOfBibitem
\bibitem[{P C Snijders, P S Johnson, P Guisinger} and Himpsel(2012){P C
  Snijders, P S Johnson, P Guisinger}, and Himpsel]{Snijders2012}
{P C Snijders, P S Johnson, P Guisinger},~S. C.~E.; Himpsel,~F.~J.
  {Spectroscopic evidence for spin-polarized edge states in graphitic Si
  nanowires}. \emph{New J.Phys} \textbf{2012}, \emph{14}, 103004\relax
\mciteBstWouldAddEndPuncttrue
\mciteSetBstMidEndSepPunct{\mcitedefaultmidpunct}
{\mcitedefaultendpunct}{\mcitedefaultseppunct}\relax
\EndOfBibitem
\bibitem[Braun \latin{et~al.}(2018)Braun, Gerstmann, and Schmidt]{Braun2018}
Braun,~C.; Gerstmann,~U.; Schmidt,~W.~G. {Spin pairing versus spin chains at
  Si(553)-Au surfaces}. \emph{Phys. Rev. B} \textbf{2018}, \emph{98},
  121402\relax
\mciteBstWouldAddEndPuncttrue
\mciteSetBstMidEndSepPunct{\mcitedefaultmidpunct}
{\mcitedefaultendpunct}{\mcitedefaultseppunct}\relax
\EndOfBibitem
\bibitem[Erwin and Himpsel(2010)Erwin, and Himpsel]{Erwin2010}
Erwin,~S.~C.; Himpsel,~F.~J. {Intrinsic magnetism at silicon surfaces.}
  \emph{Nat. Commun.} \textbf{2010}, \emph{1}, 58\relax
\mciteBstWouldAddEndPuncttrue
\mciteSetBstMidEndSepPunct{\mcitedefaultmidpunct}
{\mcitedefaultendpunct}{\mcitedefaultseppunct}\relax
\EndOfBibitem
\bibitem[Nagao(2008)]{Nagao2008}
Nagao,~T. {Characterization of atomic-level plasmonic structures by low-energy
  EELS}. \emph{Surf. Interface Anal.} \textbf{2008}, \emph{40},
  1764--1767\relax
\mciteBstWouldAddEndPuncttrue
\mciteSetBstMidEndSepPunct{\mcitedefaultmidpunct}
{\mcitedefaultendpunct}{\mcitedefaultseppunct}\relax
\EndOfBibitem
\bibitem[Rugeramigabo \latin{et~al.}(2008)Rugeramigabo, Nagao, and
  Pfn{\"{u}}r]{Rugeramigabo2008}
Rugeramigabo,~E.~P.; Nagao,~T.; Pfn{\"{u}}r,~H. {Experimental investigation of
  two-dimensional plasmons in a DySi$_2$ monolayer on Si(111)}. \emph{Phys.
  Rev. B} \textbf{2008}, \emph{78}, 1--6\relax
\mciteBstWouldAddEndPuncttrue
\mciteSetBstMidEndSepPunct{\mcitedefaultmidpunct}
{\mcitedefaultendpunct}{\mcitedefaultseppunct}\relax
\EndOfBibitem
\bibitem[Nagao \latin{et~al.}(2006)Nagao, Yaginuma, Inaoka, and
  Sakurai]{Nagao2006}
Nagao,~T.; Yaginuma,~S.; Inaoka,~T.; Sakurai,~T. {One-dimensional plasmon in an
  atomic-scale metal wire}. \emph{Phys. Rev. Lett.} \textbf{2006}, \emph{97},
  116802\relax
\mciteBstWouldAddEndPuncttrue
\mciteSetBstMidEndSepPunct{\mcitedefaultmidpunct}
{\mcitedefaultendpunct}{\mcitedefaultseppunct}\relax
\EndOfBibitem
\bibitem[Rugeramigabo \latin{et~al.}(2010)Rugeramigabo, Tegenkamp, Pfn{\"{u}}r,
  Inaoka, and Nagao]{Rugeramigabo2010}
Rugeramigabo,~E.~P.; Tegenkamp,~C.; Pfn{\"{u}}r,~H.; Inaoka,~T.; Nagao,~T.
  {One-dimensional plasmons in ultrathin metallic silicide wires of finite
  width}. \emph{Phys. Rev. B} \textbf{2010}, \emph{81}, 165407\relax
\mciteBstWouldAddEndPuncttrue
\mciteSetBstMidEndSepPunct{\mcitedefaultmidpunct}
{\mcitedefaultendpunct}{\mcitedefaultseppunct}\relax
\EndOfBibitem
\bibitem[Krieg \latin{et~al.}(2015)Krieg, Lichtenstein, Brand, Tegenkamp, and
  Pfn{\"{u}}r]{Krieg2015}
Krieg,~U.; Lichtenstein,~T.; Brand,~C.; Tegenkamp,~C.; Pfn{\"{u}}r,~H. {Origin
  of metallicity in atomic Ag wires on Si(557)}. \emph{New J. Phys.}
  \textbf{2015}, \emph{17}, 043062\relax
\mciteBstWouldAddEndPuncttrue
\mciteSetBstMidEndSepPunct{\mcitedefaultmidpunct}
{\mcitedefaultendpunct}{\mcitedefaultseppunct}\relax
\EndOfBibitem
\bibitem[Mamiyev \latin{et~al.}(2018)Mamiyev, Lichtenstein, Tegenkamp, Braun,
  Schmidt, Sanna, and Pfn\"ur]{Mamiyev2018}
Mamiyev,~Z.; Lichtenstein,~T.; Tegenkamp,~C.; Braun,~C.; Schmidt,~W.~G.;
  Sanna,~S.; Pfn\"ur,~H. Plasmon spectroscopy: Robust metallicity of Au wires
  on Si(557) upon oxidation. \emph{Phys. Rev. Materials} \textbf{2018},
  \emph{2}, 066002\relax
\mciteBstWouldAddEndPuncttrue
\mciteSetBstMidEndSepPunct{\mcitedefaultmidpunct}
{\mcitedefaultendpunct}{\mcitedefaultseppunct}\relax
\EndOfBibitem
\bibitem[Rocca(1995)]{Rocca1995}
Rocca,~M. Low-energy EELS investigation of surface electronic excitations on
  metals. \emph{Surf. Sci. Rep.} \textbf{1995}, \emph{22}, 1--71\relax
\mciteBstWouldAddEndPuncttrue
\mciteSetBstMidEndSepPunct{\mcitedefaultmidpunct}
{\mcitedefaultendpunct}{\mcitedefaultseppunct}\relax
\EndOfBibitem
\bibitem[Pitarke \latin{et~al.}(2007)Pitarke, Silkin, Chulkov, and
  Echenique]{Pitarke2007}
Pitarke,~J.~M.; Silkin,~V.~M.; Chulkov,~E.~V.; Echenique,~P.~M. {Theory of
  surface plasmons and surface-plasmon polaritons}. \emph{Reports Prog. Phys.}
  \textbf{2007}, \emph{70}, 1--87\relax
\mciteBstWouldAddEndPuncttrue
\mciteSetBstMidEndSepPunct{\mcitedefaultmidpunct}
{\mcitedefaultendpunct}{\mcitedefaultseppunct}\relax
\EndOfBibitem
\bibitem[Politano and Chiarello(2015)Politano, and Chiarello]{Politano2015}
Politano,~A.; Chiarello,~G. The influence of electron confinement, quantum size
  effects, and film morphology on the dispersion and the damping of plasmonic
  modes in Ag and Au thin films. \emph{Prog. Surf. Sci.} \textbf{2015},
  \emph{90}, 144--193\relax
\mciteBstWouldAddEndPuncttrue
\mciteSetBstMidEndSepPunct{\mcitedefaultmidpunct}
{\mcitedefaultendpunct}{\mcitedefaultseppunct}\relax
\EndOfBibitem
\bibitem[Nagao \latin{et~al.}(2010)Nagao, Han, Hoang, Wi, Pucci, Weber,
  Neubrech, Silkin, Enders, Saito, and Rana]{Nagao2010}
Nagao,~T.; Han,~G.; Hoang,~C.; Wi,~J.-S.; Pucci,~A.; Weber,~D.; Neubrech,~F.;
  Silkin,~V.~M.; Enders,~D.; Saito,~O. \latin{et~al.}  {Plasmons in nanoscale
  and atomic-scale systems}. \emph{Sci. Technol. Adv. Mater.} \textbf{2010},
  \emph{11}, 054506\relax
\mciteBstWouldAddEndPuncttrue
\mciteSetBstMidEndSepPunct{\mcitedefaultmidpunct}
{\mcitedefaultendpunct}{\mcitedefaultseppunct}\relax
\EndOfBibitem
\bibitem[Tzschoppe \latin{et~al.}(2019)Tzschoppe, Huck, H\"otzel, Butkevich,
  Mamiyev, Ulrich, Günther, Gade, and Pucci]{Michael2019}
Tzschoppe,~M.; Huck,~C.; H\"otzel,~F.; Butkevich,~A.; Mamiyev,~Z.; Ulrich,~C.;
  Günther,~B.; Gade,~L.; Pucci,~A. {How adsorbates alter the metallic behavior
  of quasi-one-dimensional electron systems of the Si(553)-Au surface}.
  \emph{J. Phys. Condens. Matter} \textbf{2019},
  \emph{doi:10.1088/1361-648X/ab0710}\relax
\mciteBstWouldAddEndPuncttrue
\mciteSetBstMidEndSepPunct{\mcitedefaultmidpunct}
{\mcitedefaultendpunct}{\mcitedefaultseppunct}\relax
\EndOfBibitem
\bibitem[Claus \latin{et~al.}(1992)Claus, B{\"{u}}ssensch{\"{u}}tt, and
  Henzler]{Claus1992}
Claus,~H.; B{\"{u}}ssensch{\"{u}}tt,~A.; Henzler,~M. {Low-energy electron
  diffraction with energy resolution}. \emph{Rev. Sci. Instrum.} \textbf{1992},
  \emph{63}, 2195\relax
\mciteBstWouldAddEndPuncttrue
\mciteSetBstMidEndSepPunct{\mcitedefaultmidpunct}
{\mcitedefaultendpunct}{\mcitedefaultseppunct}\relax
\EndOfBibitem
\bibitem[Sauerbrey(1959)]{Sauerbrey1959}
Sauerbrey,~G. {Verwendung von Schwingquarzen zur W{\"{a}}gung d{\"{u}}nner
  Schichten und zur Mikrow{\"{a}}gung}. \emph{Z. Phys.} \textbf{1959},
  \emph{155}, 206--222\relax
\mciteBstWouldAddEndPuncttrue
\mciteSetBstMidEndSepPunct{\mcitedefaultmidpunct}
{\mcitedefaultendpunct}{\mcitedefaultseppunct}\relax
\EndOfBibitem
\bibitem[Lichtenstein \latin{et~al.}(2016)Lichtenstein, Aulbach, Sch{\"{a}}fer,
  Claessen, Tegenkamp, and Pfn{\"{u}}r]{Lichtenstein2016}
Lichtenstein,~T.; Aulbach,~J.; Sch{\"{a}}fer,~J.; Claessen,~R.; Tegenkamp,~C.;
  Pfn{\"{u}}r,~H. {Two-dimensional crossover and strong coupling of plasmon
  excitations in arrays of one-dimensional atomic wires}. \emph{Phys. Rev. B}
  \textbf{2016}, \emph{93}, 161408\relax
\mciteBstWouldAddEndPuncttrue
\mciteSetBstMidEndSepPunct{\mcitedefaultmidpunct}
{\mcitedefaultendpunct}{\mcitedefaultseppunct}\relax
\EndOfBibitem
\bibitem[Ibach and Mills(1982)Ibach, and Mills]{Ibach1982}
Ibach,~H.; Mills,~D.~L. \emph{{Electron energy loss spectroscopy and surface
  vibrations}}; Academic Press, 1982\relax
\mciteBstWouldAddEndPuncttrue
\mciteSetBstMidEndSepPunct{\mcitedefaultmidpunct}
{\mcitedefaultendpunct}{\mcitedefaultseppunct}\relax
\EndOfBibitem
\bibitem[Klevenz \latin{et~al.}(2010)Klevenz, Wetzel, Trieloff, Gail, and
  Pucci]{Klevenz2010}
Klevenz,~M.; Wetzel,~S.; Trieloff,~M.; Gail,~H.-P.; Pucci,~A. Vibrational
  spectroscopy of SiO on Si (111). \emph{phys. stat. sol. b} \textbf{2010},
  \emph{247}, 2179--2184\relax
\mciteBstWouldAddEndPuncttrue
\mciteSetBstMidEndSepPunct{\mcitedefaultmidpunct}
{\mcitedefaultendpunct}{\mcitedefaultseppunct}\relax
\EndOfBibitem
\bibitem[Niu and Wang(2013)Niu, and Wang]{niu2013}
Niu,~C.-Y.; Wang,~J.-T. Adsorption and dissociation of oxygen molecules on Si
  (111)-(7$\times$ 7) surface. \emph{J. Chem. Phys.} \textbf{2013}, \emph{139},
  194709\relax
\mciteBstWouldAddEndPuncttrue
\mciteSetBstMidEndSepPunct{\mcitedefaultmidpunct}
{\mcitedefaultendpunct}{\mcitedefaultseppunct}\relax
\EndOfBibitem
\bibitem[Edler \latin{et~al.}(2017)Edler, Miccoli, St{\"{o}}ckmann,
  Pfn{\"{u}}r, Braun, Neufeld, Sanna, Schmidt, and Tegenkamp]{Edler2017}
Edler,~F.; Miccoli,~I.; St{\"{o}}ckmann,~J.~P.; Pfn{\"{u}}r,~H.; Braun,~C.;
  Neufeld,~S.; Sanna,~S.; Schmidt,~W.~G.; Tegenkamp,~C. {Tuning the
  conductivity along atomic chains by selective chemisorption}. \emph{Phys.
  Rev. B} \textbf{2017}, \emph{95}, 125409\relax
\mciteBstWouldAddEndPuncttrue
\mciteSetBstMidEndSepPunct{\mcitedefaultmidpunct}
{\mcitedefaultendpunct}{\mcitedefaultseppunct}\relax
\EndOfBibitem
\bibitem[Neubrech \latin{et~al.}(2006)Neubrech, Kolb, Lovrincic, Fahsold,
  Pucci, Aizpurua, Cornelius, Toimil-Molares, Neumann, and Karim]{Neubrech2006}
Neubrech,~F.; Kolb,~T.; Lovrincic,~R.; Fahsold,~G.; Pucci,~A.; Aizpurua,~J.;
  Cornelius,~T.~W.; Toimil-Molares,~M.~E.; Neumann,~R.; Karim,~S. Resonances of
  individual metal nanowires in the infrared. \emph{Appl. Phys. Lett.}
  \textbf{2006}, \emph{89}, 253104\relax
\mciteBstWouldAddEndPuncttrue
\mciteSetBstMidEndSepPunct{\mcitedefaultmidpunct}
{\mcitedefaultendpunct}{\mcitedefaultseppunct}\relax
\EndOfBibitem
\bibitem[Langer \latin{et~al.}(2010)Langer, Baringhaus, Pfn{\"{u}}r,
  Schumacher, and Tegenkamp]{Langer2010}
Langer,~T.; Baringhaus,~J.; Pfn{\"{u}}r,~H.; Schumacher,~H.~W.; Tegenkamp,~C.
  {Plasmon damping below the Landau regime: the role of defects in epitaxial
  graphene}. \emph{New J. Phys.} \textbf{2010}, \emph{12}, 033017\relax
\mciteBstWouldAddEndPuncttrue
\mciteSetBstMidEndSepPunct{\mcitedefaultmidpunct}
{\mcitedefaultendpunct}{\mcitedefaultseppunct}\relax
\EndOfBibitem
\end{mcitethebibliography}

%***********************************************************************
\newpage
%\Large\bf TOC graphic}
\begin{figure*}[h]
\centering{
\includegraphics[width=11.5cm]{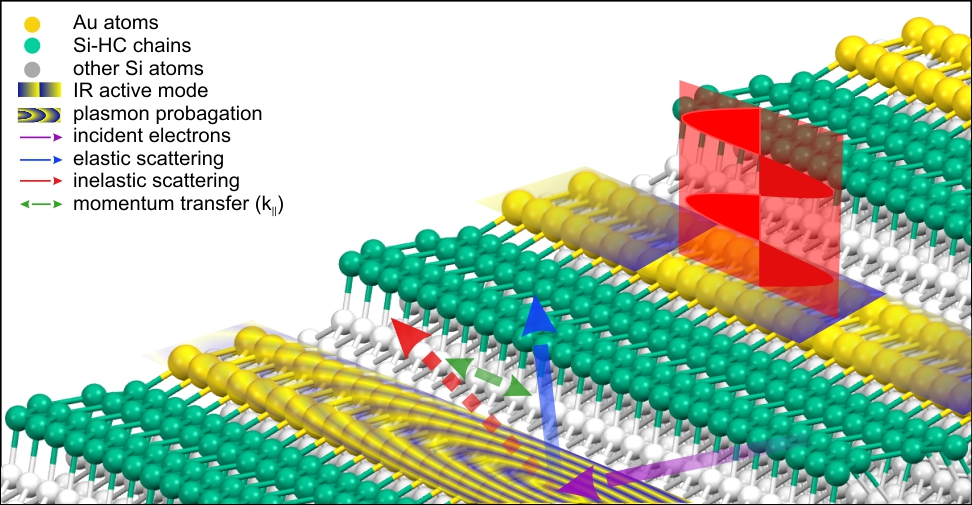} }
\end{figure*}
\end{document}